\documentclass[preprint,showpacs,amsmath,amssymb,aps,prb]{revtex4}
\usepackage{graphicx}

\begin{document}

\title{Impact of lattice strain on the tunnel magneto-resistance in Fe/Insulator/Fe and
Fe/Insulator/La$_{0.67}$Sr$_{0.33}$MnO$_{3}$ \\ magnetic tunnel junctions}

\author{A.\ Useinov$^1$\footnote{artur.useinov@csun.edu, +1-818-677-2782},
Y.\ Saeed$^2$, N.\ Singh$^2$, N.\ Useinov$^3$, U.\ Schwingenschl\"ogl$^2$\footnote{udo.schwingenschlogl@kaust.edu.sa, +966(0)544700080}}
\affiliation{$^1$Department of Physics, California State University, Northridge, California 91330, USA}
\affiliation{$^2$PSE Division, King Abdullah University of Science and Technology, Thuwal 23955, Saudi Arabia}
\affiliation{$^3$Department of Solid State Physics, Kazan Federal University, Kazan, Russia}

\date{\today}
\begin{abstract}
The objective of this work is to describe the tunnel electron current in single barrier magnetic tunnel junctions within a new approach that goes
beyond the single-band transport model. We propose a ballistic multi-channel electron transport model that can explain the influence of in-plane
lattice strain on the tunnel magnetoresistance as well as the asymmetric voltage behavior. We consider as an example single crystal magnetic Fe(110)
electrodes for Fe/Insulator/Fe and Fe/Insulator/La$_{0.67}$Sr$_{0.33}$MnO$_{3}$ tunnel junctions, where the electronic band structures of Fe and
La$_{0.67}$Sr$_{0.33}$MnO$_{3}$ are derived by \it{ab-initio} calculations.
\end{abstract}

\pacs{72.10.Fk, 73.40.Gk, 75.45.+j, 75.47.De}
\maketitle

\section{\label{sec:level1}INTRODUCTION}
One of the fast growing directions in modern magnetic electronics (spintronics) is the field of magnetic tunnel junctions (MTJs) and their applications,
for example, as basic elements in magnetic random access memories, read-heads of hard drives, and magnetic field sensors. Potential
to realize memristors and vortex oscillators creates additional incentive for future investments in this area \cite{Sp1,Sp2}.
MTJs such as FM/Insulator/FM and FM/Insulator/HM heterostructures, where FM is a ferromagnet  (like Co, Fe, CoFeB),
the insulator is ferroelectric (like BaTiO$_3$, PbTiO$_3$), and HM is a half-metal
(like La$_{0.67}$Sr$_{0.33}$MnO$_{3}$, Co$_2$MnSn), are very promising, because they combine magnetic, ferroelectric, and spin 
filtering properties. Tunnel electroresistance and tunnel magnetoresistance (TMR) effects
may coexist in these systems. The TMR arises from states of different resistance for parallel and antiparallel magnetic alignments,
while the tunnel electroresistance relies on the polarization of the ferroelectric insulator.
The insulating layer has to be thick enough to yield strong ferroelectricity, which usually rapidly disappears for decreasing thickness,
and has to be thin enough for electron tunneling. Moreover, the ferroelectric polarization in thin ferroelectric films is conjugated with
the magnitude of the lattice strain \cite{Sp3,Sp4,Sp5}. A high ferroelectric polarization is achieved by epitaxial film growth with an initially
high difference between the in-plane lattice parameters of the substrate and the deposited layers. Obviously, the electronic band structures
and transport properties of the strained FM and HM layers can be fundamentally different from those without strain.

The objective of this work is to establish the interplay between the lattice strain and the magnitude of the TMR using a multi-band approach
for the electron transport. We predict that for strained symmetric MTJs the TMR is reduced, because of changes in the electronic
band structure under strain. In general, the tunnel electroresistance in ferroelectric TJs should logarithmically increase
with strain (the ferroelectric polarization increases), as it was shown, for instance, in the works of Zhuravlev and coworkers \cite{Sp6,Sp7}.
This means there is a balanced configuration of the insulator thickness (potential barrier thickness) and strain that provides
the highest TMR and tunnel electroresistance. To calculate the tunnel current and TMR we have to go beyond the assumption of two conduction
channels (single-band model) similar to Refs.\ \cite{Sp8,Sp9,Sp10,Sp11,Sp12,Sp13,Sp14}.

Investigation of MTJs has a long history \cite{Sp15,Sp16,Sp17,Sp18}. In Ref.\ \onlinecite{Sp15} Valet and Fert have introduced basic principles
for the qualitative and quantitative interpretation of the spin polarized electron transport in magnetic multilayer structures, based on
Boltzmann-like equations. An alternative theoretical approach of electronic transport through nanocontacts with and without domain walls
between two FM electrodes has been developed in Ref.\ \onlinecite{Sp19}. This theory utilizes quasiclassical as well as quantum mechanical
ideas and is based on extended Boltzmann-like equations. Boundary conditions on the interfaces of the junction are taken into account as
a key part of the solution. The theory can be adapted to the case of ballistic transport through single barrier \cite{Sp12} and double barrier
\cite{Sp20} planar junctions. 

Using the universality of the above technique, we formulate a multi-channel (or multi-band) approach following the ideas of Ref.\ \onlinecite{Sp21}.
The tunneling conductance in MTJs can be written in terms of the averaged spin-dependent tunneling probabilities of the conduction channels for
parallel (P) and antiparallel (AP) magnetizations. According to our {\it{ab-initio}} calculations for Fe, several minority and majority spin bands
cross the Fermi level, representing different electron wave functions. We extract the dispersion relations along the tunneling
direction (perpendicular to the Fe(110) interface) from the bulk band structure. For simplicity, the insulator is considered to be homogeneous.
Our approach does not incorporate filtering effects inside the barrier, which are important in the case of MgO or for the splitting of the valence 
band in ${\rm{SrTiO}}_3$ and ${\rm{BaTiO}}_3$, for instance \cite{Sp16,Sp22,Sp23}.

\section{\label{sec:level2}THE MULTI-CHANNEL APPROACH}

\begin{figure}
\includegraphics[width=0.50\textwidth,clip]{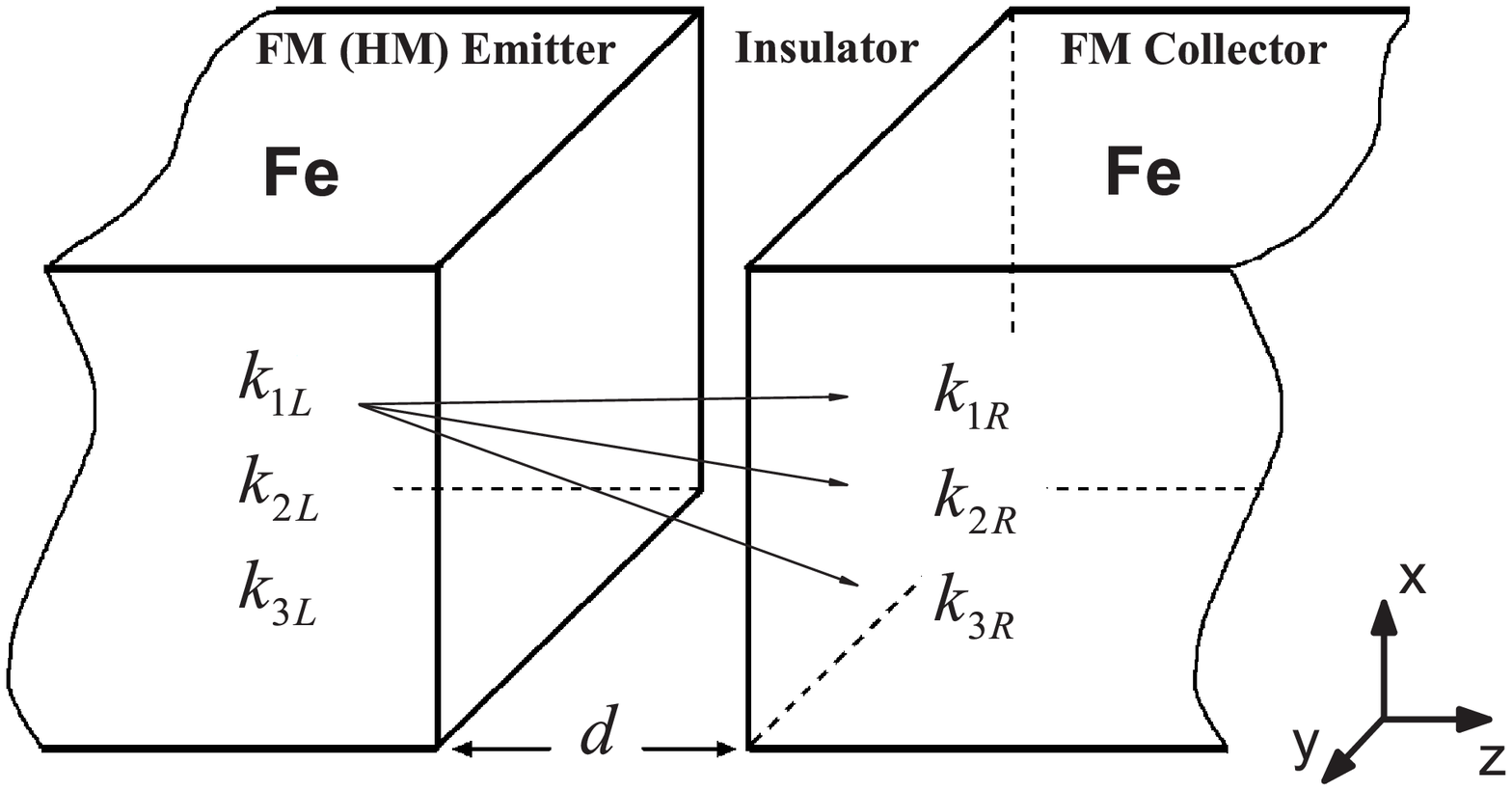}
\caption{\label{fig1} Simplified schema of the multi-channel model of a single crystal MTJ for positive bias
(electrons tunnel from left to right). 
The model assumes independent propagation channels, each being associated with a given spin and symmetry.}
\end{figure}

The ideas of the multi-channel approach are demonstrated in Fig.\ 1. In this model each propagating channel is associated with a given spin and
symmetry of the wave function. The emitter provides electrons with different Fermi vectors, which tunnel across the barrier into the states of the collector.
We employ a formula for the current density originally derived for transport through a magnetic planar junction \cite{Sp12}. For the single-band model
the current density is proportional to the integral of the product of the transmission coefficient, $D^{{\rm{P}}\left( {{\rm{AP}}} \right)}$, and the
cosine of the incidence angle of the electron trajectory, $\cos \left( {\theta _L^{ \uparrow , \downarrow } } \right)$.
The angle ${\theta _L^{ \uparrow , \downarrow } }$ is measured from the normal (transport direction) to the interface plane ($L$: left, $R$: right).
The integral is taken over $ d\Omega _L  = \sin \left( {\theta _L^{} } \right)d\theta _L^{} d\phi $:
\begin{equation}
\label{Eq.1}
J_{ \uparrow , \downarrow }^{{\rm{P}}\left( {{\rm{AP}}} \right)}  = \frac{{e^2V \left( {k_L^{ \uparrow , \downarrow } } 
\right)^2}}{{4\pi ^2 \hbar }}\left\langle {\cos \left( {\theta _L^{ \uparrow , \downarrow } } \right)D_{ \uparrow , \downarrow }^{{\rm{P}}\left( {{\rm{AP}}} \right)} } \right\rangle _{\Omega _L }. \end{equation}
Here ${k_L^{ \uparrow , \downarrow } }$ is the absolute value of the Fermi vector of the left-hand electrode and $\uparrow , \downarrow$
is the spin index. The transmission coefficient is a function of the applied bias voltage $V$, of
$\theta _L^{ \uparrow , \downarrow }  = 0...\arccos \left( {\sqrt {\left| {1 - \left( {k_R^{ \uparrow , \downarrow } /k_L^{ \uparrow , \downarrow } } \right)^2 } \right|} } \right)$,
and of $k_{L\left( R \right)}^{ \uparrow , \downarrow }$. With $x^{ \uparrow , \downarrow }  = \cos \left( {\theta _L^{ \uparrow , \downarrow } } \right)$ we can write
\begin{equation*}
\left\langle {x^{ \uparrow , \downarrow } D_{ \uparrow , \downarrow }^{{\rm{P}}\left( {{\rm{AP}}} \right)} } \right\rangle _{\Omega _L }  = \int\limits_{X^{ \uparrow , \downarrow } }^{1} {x^{ \uparrow , \downarrow } D_{ \uparrow , \downarrow }^{{\rm{P}}\left( {{\rm{AP}}} \right)} dx^{ \uparrow , \downarrow } } ,
\end{equation*}
where the lower limit  $X^{ \uparrow , \downarrow }$ for the integration arises from the conservation of the projection of the Fermi vector
in the xy-plane: $k_{\parallel}^{ \uparrow , \downarrow }  = k_L^{ \uparrow , \downarrow } \sin \left( {\theta _L^{ \uparrow , \downarrow } } \right) = k_R^{ \uparrow , \downarrow } \sin \left( {\theta _R^{ \uparrow , \downarrow } } \right) $.
It equals zero when the electrons tunnel from the left minority into the right majority conduction band and $X^{ \uparrow , \downarrow }  = \sqrt {\left| {1 - \left( {k_R^{ \uparrow , \downarrow } /k_L^{ \uparrow , \downarrow } } \right)^2 } \right|}$
when they tunnel from the left majority into the right minority conduction band.
For the multi-band approach the majority and minority bands can be both spin up and down for any magnetic configuration.

To achieve a multi-channel model (or model with multi-band tunnel relations) for single crystal junctions we redefine the current density in
Eq.\ (\ref{Eq.1}):
\begin{equation}
\label{Eq.2}
J_{ \uparrow , \downarrow }^{\text{P}\left( {\text{AP}} \right)}  = \frac{{e^2 V }}{{4\pi ^2 \hbar }}\sum\limits_{\eta  = 1}^{N } {\sum\limits_{\mu  = 1}^{M } {\left( {k_\eta ^{ \uparrow , \downarrow } } \right)^2 \left\langle {\cos \left( {\theta _\eta  } \right)D_{\eta ,\mu }^{\text{P}\left( {\text{AP}} \right)} \left( {k_\eta ^{ \uparrow , \downarrow } ,k_\mu ^{ \uparrow , \downarrow \left( { \downarrow , \uparrow } \right)} } \right)} \right\rangle _{\Omega _L } } }.
\end{equation}
Here $\eta$ and $\mu$ are the indices of the left-hand and right-hand bands, respectively, and $N$ and $M$ are the numbers of bands.
The combinations $\{\eta ,\mu \}$, see Fig.\ 1, identify the conduction relations between the bands through the barrier.
Equation (2) is valid for positive bias. The solution for negative bias is derived using symmetric
relations of the system, i.e., the collector and emitter are exchanged ($k_\eta \to k_\mu$, $k_\mu   \to k_\eta$). We assume that there is no spin flip leakage and that a conduction channel is available between any left-hand and right-hand bands
with the same spin. Otherwise the electrons are reflected back, giving rise to a resistance. Note that the lowest conductance corresponds to
the largest difference in the density of states at the Fermi level between the left and right electrodes. Regarding the transmission
coefficient for the single barrier system, the basic mathematical expressions can be found in Ref.\ \onlinecite{Sp12}, where an exact quantum mechanical
solution has been derived employing Airy functions for the tunnel barrier. 

The band structures obtained from {\it{ab-initio}} calculations for bulk Fe (space goup {\it Cmmm}) 
and La$_{0.67}$Sr$_{0.33}$MnO$_{3}$ are shown in Figs.\ 2 and 3, as derived using 
the WIEN2k package \cite{Sp24}. The exchange-correlation potential is
parametrized in the generalized gradient approximation \cite{Sp25}. For the wave function expansion inside the atomic spheres a maximum
value of the angular momentum of $\ell_{max}=12$ is employed and a plane-wave cutoff of $R_{mt}K_{max}=9$ with $G_{max}=24$ is used.
Self-consistency is assumed when the total energy variation reaches less than 10$^{-4}$ Ry. We use a mesh of $10\times10\times10$ $k$-points
for calculating the electronic structure in order to describe the ground states of the compounds with high accuracy.

\begin{figure}
\includegraphics[width=0.5\textwidth]{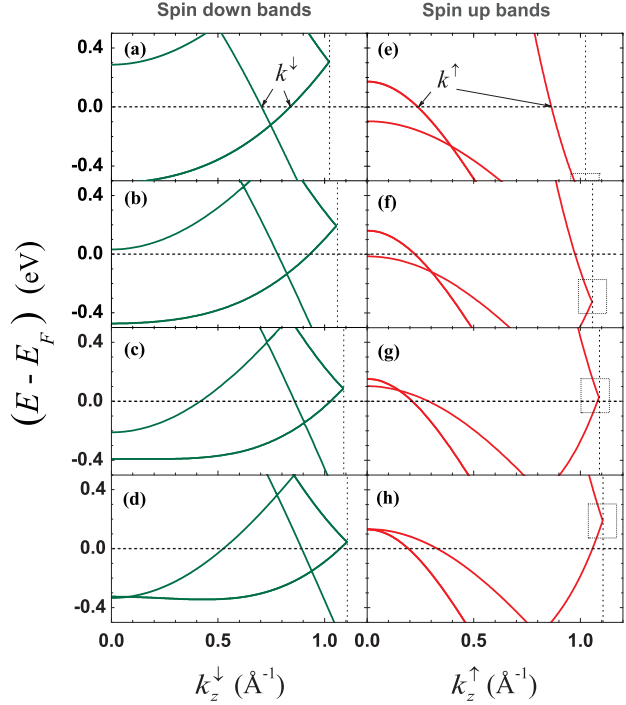}
\caption{\label{fig2} (Color online) Electronic bands for bulk Fe along the $\Gamma$-Z direction. $E-E_F=0$ corresponds to zero bias.
Data are derived for different lattice parameters, which correspond to different lattice strains. The four band structures refer to
(a),(e) $a=3.875$ \AA, $c=3.083$ \AA; (b),(f) $a=3.937$ \AA, $c=2.986$ \AA; (c),(g) $a=3.999$ \AA, 
$c=2.894$ \AA; (d),(h) $a=4.030$ \AA, $c=2.850$ \AA. The $\Gamma$ point is located at $k_z^{\uparrow ,\downarrow }=0$ 
and the Z point is shown by vertical dotted lines.}
\end{figure}

Figure 2 shows the band structure of Fe for different in-plane lattice parameters $a=3.875$\,\AA, 3.937 \AA, 3.999 \AA, and
4.030 \AA\ (bulk value), where red and green color represent the two spins. As an example, we consider the symmetric Fe/Insulator/Fe junction and demonstrate
how to collect the conducting spin channels via the applied bias $V$. The bands of the left electrode
(emitter) are the same as those of the right electrode (collector) and the Fermi energies $E_{F}^{L}=E_{F}^{R}={{E}_{F}}$ are equal at zero bias.
Horizontal dashed lines represent $E_F$, which intersects with the bands at the Fermi vectors
$k_{\eta \left(\mu\right)}^{ \uparrow , \downarrow}$. In particular, in Figs.\ \ref{fig2}(a), \ref{fig2}(b) and Figs.\ \ref{fig2}(e), \ref{fig2}(f) the system has
two $k^{\downarrow}$ and two $k^{\uparrow}$ vectors at zero bias, while in the case of Figs.\ \ref{fig2}(c), \ref{fig2}(d), \ref{fig2}(g), \ref{fig2}(h) $E_F$ is intersected by three
spin up and three spin down bands. We thus have the Fermi vector set $\{${$k_{\text{1}L\text{(}R\text{)}}^{\uparrow ,\downarrow }$, 
$k_{\text{2}L\text{(}R\text{)}}^{\uparrow ,\downarrow }$, $k_{3L\text{(}R\text{)}}^{\uparrow ,\downarrow }$}$\}$.
In the case of positive (negative) bias, by definition, $E_F$ of the left electrode shifts up (down) in energy,
while for the right electrode it shifts down (up) by the same amount. The voltage drop is $\left| E_{F}^{L}-E_{F}^{R} \right|=\left|e{V} \right|$. As a result the
Fermi vector set is changed. As an example, let us set $V=+0.8$ V with $E_{F}^{L}=0.4$ eV and $E_{F}^{R}=-0.4$ eV.
According to Figs.\ 2(a) and (e), for the left electrode this results in the Fermi vector sets $\{$0, 0, $k_{3L}^{\uparrow}\}$
and $\{k_{1L}^{\downarrow }$, $k_{2L}^{\downarrow }$, $k_{3L}^{\downarrow }\}$ and for the right electrode
in the sets $\{k_{1R}^{\uparrow}$, $k_{2R}^{\uparrow}$, $k_{3R}^{\uparrow}$ and $k_{2R}^{\downarrow}$,
$k_{3R}^{\downarrow }\}$, which generates $1\times 3=3$ channels for spin up and $3\times 2=6$ channels for
spin down, for the parallel magnetization. In contrast, $1\times 2=2$ channels for spin up and $3\times 3=9$
channels for spin down are generated in case of the antiparallel magnetization. Thus, the current can be
represented by $3\times 3=9$ channels for each spin orientation (in the general case:  $N \times M$).
When the Fermi vectors vanish we have, of course, a non-conducting channel with vanishing current density.
\begin{figure}
\includegraphics[width=0.5\textwidth]{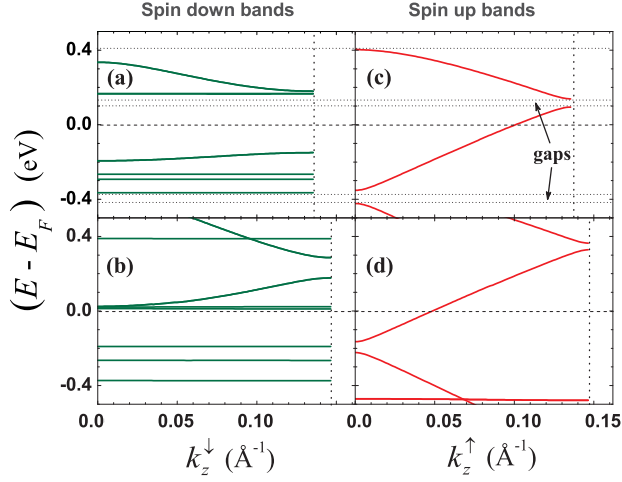}
\caption{\label{fig3} (Color online) Electronic bands for bulk La$_{0.67}$Sr$_{0.33}$MnO$_{3}$
along the $\Gamma$-Z direction. $E-E_F=0$ corresponds to zero bias. Data are derived for different lattice parameters, which correspond to
different lattice strains. The two band structures refer to (a),(c) $a=3.875$ \AA, $c=23.250$ \AA\ and (b),(d)
$a=4.030$ \AA, $c=21.496$ \AA. The $\Gamma$ point is located at $k_z^{\uparrow,\downarrow}=0$ and the Z point is shown
by vertical dotted lines.}
\end{figure}

Figure \ref{fig3} shows the band structure of La$_{0.67}$Sr$_{0.33}$MnO$_{3}$ along the $\Gamma$-Z
direction for two sets of lattice parameters: $a=3.875$ \AA, $c=23.250$ \AA\ and $a=4.030$ \AA, $c=21.496$ \AA.
The Fermi vector, transmission coefficient, and current density for each band are derived as demonstrated before.
However, some of the spin down bands are very flat with energy gaps between them, in contrast to
the spin up bands. As a function of the bias the system therefore switches between a HM
and FM. However, there are also energies at which neither spin up nor spin down states exist. 

\section{TUNNEL MAGNETORESISTANCE UNDER STRAIN}

Physical parameters that characterize the properties of MTJs are the total tunnel current density
$J^{\text{P}\left( \text{AP} \right)}={{\left( {{J}_{\uparrow }}+{{J}_{\downarrow }} \right)}^{\text{P}\left( \text{AP} \right)}}$, 
the $\text{TMR}={\left( {{J}^{\text{P}}}-{{J}^{\text{AP}}} \right)}/{{{J}^{\text{AP}}}}\times100\%$,
the normalized ${\rm{TMR}}_{\rm{n}}  = \left( {J^{\rm{P}}  - J^{{\rm{AP}}} } \right)/J^{{\rm{AP}}}  
\times {\rm{TMR}}_{}^{ - 1} \left( {V = 0} \right)$, and the output voltage ${{V}_\text{out}}={{V}\left( {{J}^{\text{P}}}-{{J}^{\text{AP}}} \right)}/{{{J}^{\text{AP}}}}$,
which can be obtained from free-electron \cite{Sp11,Sp26} or tight-binding
\cite{Sp13,Sp14} models. However, unfortunately these models do not reproduce the
experimental effect of strain on the charge transport characteristics.
A single-band approach is sufficient to model the TMR in amorphous sputtered MTJs \cite{Sp27}
and can satisfactorily describe the ${\rm{TMR}}_{\rm{n}}$ and ${{V}_\text{out}}$ of epitaxial single and
double barrier FeCoB/MgO junctions \cite{Sp28}. In our case we have to go beyond parabolic
dispersions and the single-band model, however, keeping the simplicity of the approach.
\begin{figure}
\includegraphics[width=0.50\textwidth]{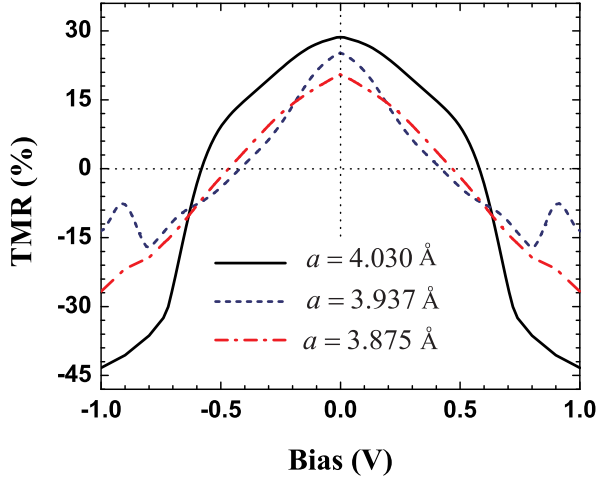}
\caption{\label{fig4} (Color online) TMR versus applied voltage for Fe/Insulator/Fe MTJs with
the lattice parameters: $a=3.875$ \AA, 3.937 \AA, and 4.030 \AA. The barrier parameters are
$d=1.8$ nm and $U_B=2.8$ eV.}
\end{figure}
For the Fermi vectors derived above as well as for typical parameters of an Al$_2$O$_3$
tunnel barrier, TMR results derived by Eq.~(2) are shown in Figs.~\ref{fig4} to \ref{fig6}.
The barrier thickness is set to $d=1.8$ nm, the barrier height above $E_F$
to $U_B=2.8$ eV, and the effective mass to $m_B=0.25$ \cite{Sp29}. In our
calculations for metals the effective mass is equal to the free electron mass.

Figure \ref{fig4} presents the TMR as function of the bias for different lattice parameters,
showing that the TMR, in general, behaves non-monotonically. For unstrained Fe
($a=4.030$ \AA) a decreasing in-plane lattice parameter
(increasing strain) leads to a lower TMR. Figure \ref{fig5} gives the TMR as a
function of the lattice parameter for 0.1~mV and 0.1~V bias. Interestingly,
we observe deviations from a linear behavior: For almost zero bias the TMR increases
up to 31.1\% for $a =3.999$ \AA, 28.6\% for $a=4.030$ \AA, and 27.3\% for
$a =3.968$ \AA. This behavior is related to modifications in the reflection
of the majority states at the Z point, where the Fermi vector achieves its maximal
magnitude (Fig.\ 2, dashed rectangles). Note that these states give the main contribution
to the tunnel current. The observed differences for different in-plane lattice parameters
are explained by variations of the band structure. The dashed rectangles in Figs.\ 2(e-h)
demonstrate the bands near the Z point. For $a =3.999$ \AA, see
Fig.\ 2(g), the majority band intersects the Fermi level at the Z point, favoring
$J^{\rm{P}}$ over $J^{\rm{AP}}$, in contrast to the other lattice parameters. The maximal
TMR value close to zero bias is in good agreement with the results of Yuasa
and coworkers for Fe(110)/Al$_2$O$_3$/Fe$_{50}$Co$_{50}$, see
Fig.\ 3(b) in Ref.\ \onlinecite{Sp29}, and of Hauch and coworkers for Fe(110)/MgO(111)/Fe(110),
28\% at $T=300$ K \cite{Sp30}.

\begin{figure}
\includegraphics[width=0.50\textwidth]{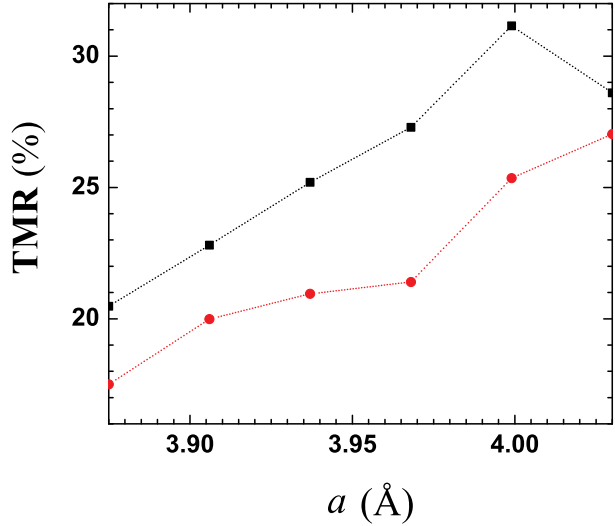}
\caption{\label{fig5} (Color online) TMR as function of the lattice parameter $a$ for
Fe/Insulator/Fe MTJs. Black and red color refer to biases of 0.1 mV and 0.1 V, respectively.}
\end{figure}

In the case of the Fe/Insulator/La$_{0.67}$Sr$_{0.33}$MnO$_{3}$ MTJ our model gives a positive TMR
for $V>0.11$ V as well as a negative TMR below, see Fig.\ \ref{fig6}. The TMR curves are
qualitatively similar to those obtained experimentally for Co/SrTiO$_3$/La$_{0.7}$Sr$_{0.3}$MnO$_3$ \cite{Sp31}
and agree with the room temperature TMR in ${\rm{Fe/MgO/Co}}_2 {\rm{MnSn}}$ \cite{Sp32} (about $-5$\% at
a bias of 0.1 mV). However, according to these authors the TMR is suppressed
in the voltage range $|V|\ge0.5$\,V, which is probably related to enhanced spin scattering for high bias.
TMR curves are given in Fig.\ 6 for the in-plane lattice parameters $a=4.030$ \AA\
and $a=3.875$ \AA, where the latter corresponds to unstrained La$_{0.67}$Sr$_{0.33}$MnO$_{3}$.
For positive bias the magnitude of the TMR decreases with the Fe lattice strain, whereas for negative
bias the situation is reversed. For the circled points in Fig.\ 6, where the TMR goes to zero,
both spin channels are closed, compare the energy gaps in Fig.\ 3, because of $J^{\rm{P}} = J^{{\rm{AP}}} = 0$.
There are other points where the TMR is zero as $J^{\rm{P}} = J^{\rm{AP}}$.
Variation of the effective mass in the tunnel barrier leads to a weak response of the TMR
in symmetric (1.5\% decrease) and a strong response in asymmetric (14\% increase) junctions,
for all lattice parameters close to zero bias, $m_B=1$.

\begin{figure}
\includegraphics[width=0.50\textwidth]{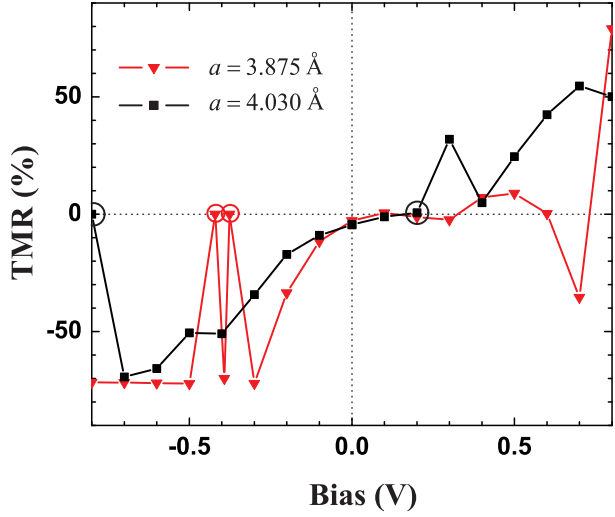}
\caption{\label{fig6}(Color online) TMR versus applied bias for the Fe/Insulator/La$_{0.67}$Sr$_{0.33}$MnO$_{3}$ MTJ.
The barrier parameters are $d=1.8$ nm, $U_B=2.8$ eV, and $m_B=0.25$.}
\end{figure}

\section{CONCLUSION}

We have extended an established quasi-classical ballistic transport model to multi-channel
conductance, which has enabled us to investigate the role of the electronic band structure
and the effect of strain on the transport properties of single crystal Fe/Insulator/Fe
and Fe/Insulator/La$_{0.67}$Sr$_{0.33}$MnO$_{3}$ MTJs. Our approach takes into account
all bands of the FM and HM along the $\Gamma$-Z direction (direction of
tunneling). We have demonstrated for typical parameters of an Al$_2$O$_3$ tunnel barrier
a maximal TMR of 31.1\% for the Fe/Insulator/Fe MTJ, which is in good agreement with the
experiment. A negative TMR of 5\% is found for the Fe/Insulator/La$_{0.67}$Sr$_{0.33}$MnO$_{3}$
MTJ close to zero bias, where the dependence on the bias reproduces experimental findings. The
developed technique thus has demonstrated great potential for further studies on transport
properties (including the spin transfer torque) in simple and magnetic TJs.

Strain effects on the TMR have been explored theoretically for the first time by a multi-band
approach. For the Fe/Insulator/Fe MTJ it turnes out that for small bias the TMR decreases
linearly with the in-plane strain at the interface, whereas in the case of the
Fe/Insulator/La$_{0.67}$Sr$_{0.33}$MnO$_{3}$ MTJ the strain effects strongly depend on the
sign of the applied bias. For positive bias it is positive and maximal for 
unstrained Fe, while for negative bias it is negative and the amplitude increases with
strain and bias. The observed relations between the strain and the TMR are explained
by variations of the band structure. We have demonstrated that in-plane strain can
increase and decrease the TMR and therefore makes it possible to obtain optimal
regimes for MTJ applications.

\end{document}